\documentclass[11pt,a4paper]{article}
\usepackage[hyperref]{naaclhlt2018}
\usepackage{times}
\usepackage{latexsym}
\RequirePackage{amsmath,amsthm,amsfonts,amssymb}
\usepackage{amsmath}
\usepackage{booktabs}
\usepackage{multirow}
\usepackage{graphicx}
\usepackage{setspace, amssymb}
\usepackage{comment}

\usepackage{url}
\aclfinalcopy % Uncomment this line for all SemEval submissions

\title{Cross-Lingual Cross-Corpus Speech Emotion Recognition}
\begin{document}

\author{Shivali Goel \hspace{65pt} Homayoon Beigi \\
  Department of Computer Science, Columbia University \\
  {\tt \{sg3629, hb87\}@columbia.edu}
  %\\\AND Homayoon Beigi \\
  %Department of Computer Science, Columbia University \\
  %{\tt hb87@columbia.edu}
  }
\maketitle
\begin{abstract}
The majority of existing speech emotion recognition models are trained and evaluated on a single corpus and a single language setting. These systems do not perform as well when applied in a cross-corpus and cross-language scenario. This paper presents results for speech emotion recognition for 4 languages in both single corpus and cross corpus setting. Additionally, since multi-task learning (MTL) with gender, naturalness and arousal as auxiliary tasks has shown to enhance the generalisation capabilities of the emotion models, this paper introduces language ID as another auxiliary task in MTL framework to explore the role of spoken language on emotion recognition which has not been studied yet.

\end{abstract}

%{\let\thefootnote\relax\footnote{{* Equal contribution.}}}

\section{Introduction}

Speech conveys human emotions most naturally. In recent years there has been an increased research interest in speech emotion recognition domain. The first step in a typical SER system is extracting linguistic and acoustic features from speech signal. Some para-linguistic studies find Low-Level Descriptor (LLD) features of the speech signal to be most relevant to studying emotions in speech. These features include frequency related parameters like pitch and jitter, energy parameters like shimmer and loudness, spectral parameters like alpha ratio and other parameters that convey cepstral and dynamic information. Feature extraction is followed with a classification task to predict the emotions of the speaker.

Data scarcity or lack of free speech corpus is a problem for research in speech domain in general. This also means that there are even fewer resources for studying emotion in speech. For those that are available are dissimilar in terms of the spoken language, type of emotion (i.e. naturalistic, elicited, or acted) and labelling scheme (i.e. dimensional or categorical). 

Across various studies involving SER we observe that performance of model depends heavily on whether training and testing is performed from the same corpus or not. Performance is best when focus is on a single corpus at a time, without considering the performance of model in cross-language and cross-corpus scenarios. In this work, we work with diverse SER datasets i.e. tackle the problem in both cross-language and cross-corpus setting. We use transfer learning across SER datasets and investigate the effects of language spoken on  the accuracy of the  emotion  recognition  system using our Multi-Task Learning framework.

The paper is organized as follows: Section 2 reviewed related work on SER, cross-lingual and cross-corpus SER and  the  recent studies  on  role of language identification in speech emotion  recognition  system, Section  3  describes  the  datasets  that  have  been  used,  Section  4  presents detailed descriptions of three types of SER experiments we conduct in this paper. In  Section 5, we present our results and evaluations of our models. Section 6 presents some additional experiments to draw a direct comparison with previously published research. Finally, we discuss future work and conclude the paper.

\begin{comment}
Transfer learning models like ULMFiT, BERT and GPT \cite{Ulmfit, bert, gpt} have shown their effectiveness on several NLP tasks. There are a few studies involving transfer learning for cross-corpus emotion recognition but it still has not been thoroughly explored. In recent years researchers have also been looking into the use of multi-modal features for emotion recognition, however there is no work that covers both cross-corpus and multi-modal aspect of SER. 

Second part of this study tries to address the above challenge. We propose a novel SER transfer learning approach for cross-corpus emotion recognition. Our approach uses both textual and speech features. For each dataset, we compute MFCC as speech features and word embeddings as textual features. To use different language datasets for training a single classifier we embed the word embeddings in different languages into a shared vector space by using cross lingual word embeddings. We also only consider 6 emotions common across all datasets. By this approach we essentially create a unified dataset from various different emotion dataset and train a single deep learning based emotion classifier to improve emotion detection accuracy. Datasets used are listed in section 3 and detailed approach is discussed in section 4 of the paper. 
\end{comment}

\begin{table*}[!htb]
{\small
\centering
\begin{tabular}{|l|l|l|l|l|}
\hline
\textbf{Dataset} & \textbf{Language} & \textbf{Utterances} & \textbf{\begin{tabular}[c]{@{}l@{}}\#Emotion \\ categories\end{tabular}} & \textbf{Emotion labels} \\ \hline
EMO-DB & German & 494 & 7 & Anger, Sadness, Fear, Disgust, Boredom, Neutral, Happiness \\ \hline
SAVEE & English & 480 & 7 & Anger, Sadness, Fear, Disgust, Neutral, Happiness, Surprise \\ \hline
EMOVO & Italian & 588 & 7 & Anger, Sadness, Fear, Disgust, Neutral, Joy, Surprise \\ \hline
MASC & Chinese & 25636 & 5 & Anger, Sadness, Panic, Neutral, Elation \\ \hline
IEMOCAP & English & scripted: 5255 turns; & 9 & Anger, Happiness, Excitement, Sadness, Frustration, Fear,\\ & & spontaneous: 4784 turns & & Surprise, Other and Neutral \\ \hline
\end{tabular}
}
\caption{Datasets used for various SER experiments.}
\end{table*}

\begin{table*}[!htb]
\begin{tabular}{|l|l|l|l|l|l|l|}
\hline
\multirow{2}{*}{Feature Set} & \multicolumn{1}{c|}{\multirow{2}{*}{Classifier}} & \multicolumn{5}{c|}{Dataset} \\ \cline{3-7} 
 & \multicolumn{1}{c|}{} & EMODB & EMOVO & SAVEE & IEMOCAP & MANDARIN \\ \hline
MFCC & LSTM & 44.19 & 30 & 35.56 & 50.48 & 41.64 \\ \hline
\multirow{3}{*}{IS09 Emotion} & Logistic Regression & 85 & 38 & 46 & 58 & 49 \\ \cline{2-7} 
 & SVC & 88.37 & 35.71 & 55.55 & 61.20 & 58 \\ \cline{2-7} 
 & LSTM & 86.05 & 27.14 & 55.56 & 55.09 & 50.40 \\ \hline
\end{tabular}
\caption{SER performance for each of the 5 datasets using  different feature sets and classifiers}
\end{table*}

\begin{table*}[]
\begin{tabular}{|l|l|l|l|l|l|}
\hline
\multirow{2}{*}{Feature Set} & \multirow{2}{*}{Classifier} & \multirow{2}{*}{Train} & \multicolumn{3}{l|}{Test} \\ \cline{4-6} 
 &  &  & EMODB & EMOVO & SAVEE \\ \hline
\multirow{2}{*}{IS09 Emotion} & \multirow{2}{*}{LSTM} & Train on IEMOCAP & 0.4651 & 0.3571 & 0.4555 \\ \cline{3-6} 
 &  & Fine-tune on smaller dataset & 0.8372 & 0.3142 & 0.5555 \\ \hline
\end{tabular}
\caption{Transfer learning for small datasets. Row 1: Training on large English corpus, testing on test sets of small corpses. Row 2: Fine-tune base English model on say EMODB train set and test on EMODB test set}
\end{table*}

\begin{table*}[]
\begin{tabular}{|l|l|l|l|l|l|l|}
\hline
\multirow{2}{*}{\begin{tabular}[c]{@{}l@{}}Feature\\ Set\end{tabular}} & \multirow{2}{*}{Classifier} & \multicolumn{5}{l|}{Test} \\ \cline{3-7} 
 &  & EMODB & EMOVO & SAVEE & IEMOCAP & MANDARIN \\ \hline
\multirow{2}{*}{MFCC} & LSTM (only predict emotion) & 58.14 & 21.43 & 34.44 & 50.80 & 43.37 \\ \cline{2-7} 
 & \begin{tabular}[c]{@{}l@{}}Multi-task LSTM (predict\\ both emotion and language ID)\end{tabular} & 53.48 & 28.00 & 33.30 & 50.69 & 43.10 \\ \hline
\end{tabular}
\caption{Multitask-learning. Table only shows accuracy scores for emotion recognition. Model always predicted language ID with very high accuracy($>$97\%).}
\end{table*}

\section{Related Work}

Over the last two decades there have been considerable research work on speech emotion recognition. However, all these differ in terms of the training corpora, test conditions, evaluation strategies and more which create difficulty in reproducing exact results.  In \cite{inproceedings123}, the authors give an overview of types of features, classifiers and emotional speech databases used in various SER research.

Speech emotion recognition has evolved over time with regards to both the type of features and models used for classifiers. Different types of features that can be used can involve simple features like pitch and intensity \cite{Rychlicki-Kicior_K._Multipitch_2014, article123}. Some studies use low-level descriptor features(LLDs) like jitter, shimmer, HNR and spectral/cepstral parameters like alpha ratio \cite{article31,10.1007/978-3-540-74889-2_13}. Other features include rhythm and sentence duration \cite{5209158} and non-uniform perceptual linear predictive (UN- PLP) features \cite{5401296}. Sometimes, linear predictive cepstral coefficients(LPCCs) \cite{5170314} are used in conjunction with mel-frequency cepstral coefficients (MFCCs).

There have been studies on SER in languages other than english. For example, \cite{7816929} propose a deep learning model consisting of stacked auto-encoders and deep belief networks for SER on the famous German dataset EMODB. \cite{87f0a11a09104176be8bee8177316fef} were the first to study SER work on the GEES, a Serbian emotional speech corpus. The authors developed a multistage strategy with SVMs for emotion recognition on a single dataset. 

Relatively fewer studies address the problem of cross-language and cross-corpus speech emotion recognition. \cite{SCHULLER20111062,Schuller:2010:CAE:1936323.1936497}. Recent work by \cite{DBLP:journals/corr/abs-1812-10411, Latif2019UnsupervisedAD} studies SER for languages belonging to different language families like Urdu vs. Italian or German. Other work involving cross-language emotion recognition includes \cite{7949505} which studies speech emotion recognition for for mandarin language vs. western languages like German and Danish. \cite{Albornoz:2017:ERN:3068941.3068956} developed an ensemble SVM for emotion detection with a focus on emotion recognition in unseen languages.%The authors focused on gender-specific speech emotion recognition and achieved the classification rates higher than the chance level but less than baseline accuracy. 

Although there are a lot of psychological case studies on the effect of language and culture in SER, there are very few computational linguistic studies in the same domain. In \cite{7575033}, the authors support the fact that SER is language independent, however also reveal that there are language specific differences in emotion recognition in which English shows a higher recognition rate compared to Malay and Mandarin. In \cite{10.1371/journal.pone.0220386} the authors proposed two-pass method based on language identification and then emotion recognition. It showed significant improvement in performance. They used English IEMOCAP, the German Emo-DB, and a Japanese corpus to recognize four emotions based on the proposed two-pass method. 

In \cite{inproceedings12345}, the authors also use language identification to enhance cross-lingual SER. They concluded that in order to recognize the emotions of a speaker whose language is unknown, it is beneficial to use a language identifier followed by model selection instead of using a model which is trained based on all available languages. This work is to the best of our knowledge the first work that jointly tries to learn the language and emotion in speech. 

%\paragraph{Speech Emotion Recognition.} 

%\paragraph{Cross-Corpus and Cross-Lingual SER} 

%\paragraph{Multi-modal SER} 

\section{Datasets}

\paragraph{EMO-DB} 
This dataset was introduced by \cite{Burkhardt05adatabase}. Language of recordings is German and consists of acted speech with 7 categorical labels. The semantic content in this data is pre-defined in 10 emotionally neutral German short sentences. It contains 494 emotionally labeled phrases collected from 5 male and 5 female actors in age range of 21-35 years.

\paragraph{SAVEE}
Surrey Audio-Visual Expressed Emotion (SAVEE) database \cite{dataset} is a famous acted-speech multimodal corpus. It consists of 480 British English utterances from 4 male actors in 7 different emotion categories. The text material consisted of 15 TIMIT \cite{timit} sentences per emotion: 3 common, 2 emotion-specific and 10 generic sentences that were different for each emotion and phonetically-balanced.

\paragraph{EMOVO} 
This \cite{costantini-etal-2014-emovo} is an Italian language acted speech emotional corpus that contains recordings of 6 actors who acted 14 emotionally neutral short sentences sentences to simulate 7 emotional states. It consists of 588 utterances and annotated by two different groups of 24 annotators. 

\paragraph{MASC: Mandarin Affective Speech Corpus}
This is an Mandarin language acted speech emotional corpus that consist of 68 speakers (23 females, 45 males) each reading out read that consisted of five phrases, fifteen sentences and two paragraphs to simulate 5 emotional states. Altogether this database \cite{4013501} contains 25,636 utterances.

\paragraph{IEMOCAP: The Interactive Emotional Dyadic Motion Capture} IEMOCAP database \cite{articleIEMOCAP} is an English language multi-modal emotional speech database. It contains approximately 12 hours of audiovisual data, including video, speech, motion capture of face, text transcriptions. It consists of dyadic sessions where actors perform improvisations or scripted scenarios, specifically selected to elicit emotional expressions. It has categorical labels, such as anger, happiness, sadness, neutrality, as well as dimensional labels such as valence, activation and dominance. 

\section{Experiments}
\begin{figure}[h]
\centering
\includegraphics[width=6cm]{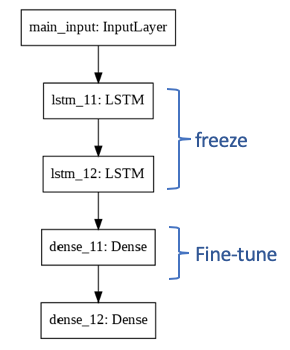}
\caption{Transfer learning for small datasets}
\end{figure}

\begin{figure}[h]
\centering
\includegraphics[width=7cm]{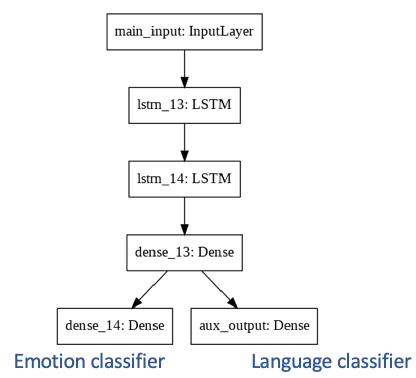}
\caption{Multi-task learning for learning emotion and language ID simultaneously}
\end{figure}

\subsection{SER on Individual Datasets}
The first set of experiments focused on performing speech emotion recognition for the 5 datasets individually. We perform a 5-way classification  by choosing 5 emotions common in all datasets i.e. happy, sad, fear, anger and neutral. For each dataset, we experiment with different types of features and classifiers. To generate Mel-frequency Cepstral Coefficients (MFCC) features we used the Kaldi-toolkit. We created spk2utt, utt2spk and wav.scp files for each dataset and generated MFCC features in .ark format. We leveraged kaldiio python library to convert .ark files to numpy arrays. Apart from MFCC's we also computed pitch features using the same toolkit. We keep a maximum of 120 frames of the input, and zero padded the extra signal for short utterances and clipped the extra signal for longer utterances to end up with (120,13) feature vector for each utterance. 

To compare emotion classification performance using MFCC's as input features we also tried a different feature set i.e.  IS09 emotion feature set \cite{IS09feat} which has in previous research shown good performance on SER tasks. The IS09 feature set contains 384 features that result from a systematic combination of 16 Low-Level Descriptors (LLDs) and corresponding first order delta coefficients with 12 functionals. The 16 LLDs consist of zero-crossing-rate (ZCR), root mean square (RMS) frame energy, pitch frequency (normalized to 500 Hz), harmonics-to-noise ratio (HNR) by autocorrelation function, and mel-frequency cepstral coefficients (MFCC) 1–12 (in full accordance to HTK-based computation). The 12 functionals used are mean, standard deviation, kurtosis, skewness, minimum, maximum, relative position, range, and offset and slope of linear regression of segment contours, as well as its two regression coefficients with their mean square error (MSE) applied on a chunk. To get these features we had to install OpenSmile toolkit. Script to get these features after installation is included in code submitted (refer IS09 directory).  

Once we had our input features ready we created test datasets from each of the 5 datasets by leaving one speaker out for small datasets (EMOVO, EMODB, SAVEE) and 2 speakers out for the larger datasets (IEMOCAP, MASC). Thus, for all corpora, the speakers in the test sets do not appear in the training set. We then performed SER using both classical machine learning and deep learning models. We used Support Vector one-vs-rest classifier and  Logistic Regression Classifier for classical ML models and a stacked LSTM model for the deep learning based classifier. The LSTM network comprised of 2 hidden layers with 128 LSTM cells, followed by a dense layer of size 5 with softmax activation. 

We present a comparative study across all datasets, feature sets and classifiers in table 2.

\subsection{SER using Transfer learning for small sized datasets}
In the next step of experiments we tried to improve on the results we got for individual datasets by trying to leverage the technique of transfer learning. While we had relatively large support for languages like English and Chinese, speech emotion datasets for other languages like Italian and German were very small i.e. only had a total of around 500 labeled utterances. Such small amount of training data is not sufficient specially when training a deep learning based model.

We used the same LSTM classifier as detailed in section 4.1. with an additional dense layer before the final dense layer with softmax. We train this base model using the large IEMOCAP English dataset. We then freeze the weights of LSTM layers i.e. only trainable weights in the classifier remain those of the penultimate dense layer. We fine tune the weights of this layer using the small datasets(eg. SAVEE, EMODB, EMOVO) and test performance on the same test sets we created in section 4.1. 

Table 3 shows the results of transfer learning experiments.

\subsection{Multitask learning for SER}
Last set of experiments focus on studying the role of language being spoken on emotion recognition. Due to the lack of adequately sized emotion corpus in many languages, researchers have previously tried training emotion recognition models on cross-corpus data i.e. training with data in one or more language and testing on another. This approach sounds valid only if we consider that expression of emotion is same in all languages i.e. no matter which language you speak, the way you convey your happiness, anger, sadness etc will remain the same. One example can be that low pitch signals are generally associated with sadness and high pitch and amplitude with anger. If expression of emotion is indeed language agnostic we could train emotion recognition models with high resource languages and use the same models for low resource languages.

To verify this hypothesis, we came up with a multi-task framework that jointly learns to predict emotion and the language in which the emotion is being expressed. The framework is illustrated in figure 2. The parameters of the LSTM model remain the same as mentioned in section 4.1. The SER performance of using training data from all languages and training a single classifier(same as shown in figure 1) vs. using training data from all languages in a multi-task setting is mentioned in table 4.

\section{Results and Analysis}

We will discuss the results of each experiment in detail in this section:

\begin{enumerate}
\item For SER experiments on individual dataset we see from Table 2 that SVC classifier with IS09 input features gave the best performance for four out of 5 datasets. We also note a huge difference in accuracy scores when using the same LSTM classifier and only changing the input features i.e. MFCC and IS09. LSTM model with IS09 input features gives better emotion recognition performance for four out of 5 datasets. These experiments suggest the superiority of IS09 features as compared to MFCC's for SER tasks. 

\item As expected the second set of experiments show that transfer learning is beneficial for SER task for small datasets. In table 3 we observe that training on IEMOCAP and then fine-tuning on train set of small dataset improves performance for german dataset EMODB and smaller english dataset SAVEE. However, we also note a small drop in performance for Italian dataset EMOVO. 

\item Results in table 4 do not show improvement with using language as an auxiliary task in speech emotion recognition. While a improvement would have suggested that language spoken does affect the way people express emotions in speech, the current results are more suggestive of the fact that emotion in speech are universal i.e. language agnostic. People speaking different languages express emotions in the same way and SER models could be jointly trained across various SER corpus we have for different languages. 

\end{enumerate}

\section{Comparison with Previous Research}

In this section we present comparative study of two previous research papers with our work. We keep this report in a separate section because in order to give a direct comparison with these two papers we had to follow their train-test split, number of emotion classes etc. 

\begin{enumerate}
\begin{table}[h]
\small
\begin{tabular}{|l|l|l|l|l|}
\hline
\multirow{2}{*}{} & Train & \multicolumn{3}{l|}{Test} \\ \cline{2-5} 
 & IEMOCAP & EMOVO & SAVEE & EMODB \\ \hline
Parry et al. & 51.45 & 33.33 & 33.33 & 41.99 \\ \hline
\begin{tabular}[c]{@{}l@{}}Ours\\ (IS09 + SVC)\end{tabular} & 61.00 & 32.00 & 51.00 & 65.00 \\ \hline
\begin{tabular}[c]{@{}l@{}}Ours\\ (IS09 + LSTM)\end{tabular} & 55.20 & 31.43 & 43.33 & 46.51 \\ \hline
\end{tabular}
\caption{Comparative results with Parry et al.}
\end{table}

\item {In Analysis of Deep Learning Architectures for Cross-corpus Speech Emotion Recognition \cite{Parry2019}, the authors discuss cross-corpus training using 6 datasets. In one of their experiments, they report performance on test set of each corpus for models trained only on IEMOCAP dataset. When we perform the same experiment i.e. train our model only on IEMOCAP and test on other datasets using IS09 as input features and SVC classifier, we observe better results even while performing a 5 way classification task as compared to their 4 way classification. Results are shown in Table 5.}

\item{In multi modal emotion recognition on IEMOCAP with neural networks \cite{articlehooma}, the authors present three deep learning based speech emotion recognition models. We follow the exact same data pre-processing steps for obtaining same train-test split. We also use the same LSTM model as their best performing model to verify we get the same result i.e. accuracy of 55.65\%. However, we could improve this performance to 56.45\% by using IS09 features for input and a simple SVC classifier. This experiment suggested we could get equal or better performance in much less training time with classical machine learning models given the right input features as compared to sophisticated deep learning classifiers.}

\end{enumerate}

\section{Future Work}
In future we would like to experiment with more architectures and feature sets. We would also like to extend this study to include other languages, specially low resource languages. Since all datasets in this study were acted speech, another interesting study would be to note the differences that arise when dealing with natural speech.

\section{Conclusion}
Some of the main conclusions that can be drawn from this study are that classical machine learning models may perform as well as deep learning models for SER tasks given we choose the right input features. IS09 features consistently perform well for SER tasks across datasets in different languages. Transfer learning proved to be an effective technique for performing SER for small datasets and multi-task learning experiments shed light on the language agnostic nature of speech emotion recognition task.

\bibliography{naaclhlt2018}
\bibliographystyle{acl_natbib}

\end{document}